\documentstyle[aps,prd,preprint]{revtex}
\begin{document}
\draft
\preprint{\vbox{ \hbox{SNUTP 96/043} \hbox{hep-th/9605176} }}
\title{Back Reaction and Graceful Exit in String Inflationary Cosmology}
\author{Soo-Jong Rey}
\address{
Physics Department \& Center for Theoretical Physics\\
Seoul National University, Seoul 151-752 KOREA\\
}
%\date{}
\maketitle

\begin{abstract}
\noindent
Classical string cosmology consists of two branches related to each other by 
scale-factor duality: a super-inflation branch and a 
Friedmann-Robertson-Walker (FRW) branch. 
Curvature and string coupling singularity separates the two branches, 
hence posing `graceful exit problem' to super-inflationary string cosmology. 
In an exactly soluble two-dimensional compactification model it is shown that 
quantum back reaction retards curvature and string coupling growth and 
connects the super-inflation branch to the FRW branch without encountering
a singularity. This may offer an attractive solution to the `graceful exit
problem' in string inflationary cosmology.
\end{abstract}
%\narrowtext
In the Standard Model of big bang cosmology horizon and flatness problems 
constitute two notorious naturalness problems. Inflationary 
cosmology~\cite{infl} was 
invented to solve these problems through a long period of accelerated 
expansion $\dot a (t) > 0$, $\ddot a(t) > 0$. Three different types of 
inflation satisfy these conditions: de Sitter $a(t) \sim e^{Ht}$, power-law 
$a(t) \sim t^p (p > 1)$ and super-inflation $a(t) \sim (-t)^p (p < 0)$. 
Among them super-inflation is regarded as the most attractive: 
super-inflation is driven by kinetic energy, hence, largely model-independent 
and free from fine-tuning problem typically present in potential energy driven 
de Sitter or power-law inflation models.

Veneziano~\cite{vene} has first pointed out that, 
in classical string cosmology, thanks to the scale-factory duality the 
super-inflation arises naturally in addition to decelerating expansion of 
Friedmann-Robertson-Walker (FRW) type. One might then hope that superstring 
theory gives rise to a successful big bang cosmology starting from a 
super-inflationary branch and then smoothly evolve into a FRW-type branch. 
It has been noted, however, that such evolution cannot be realized since 
the super-inflation phase ends with divergent curvature and string coupling 
and cannot be continued to regular FRW phase. Because of this string
inflationary cosmology is afflicted by a `graceful exit 
problem'~\cite{gracefulexit}.

The fact that both curvature and string coupling grow arbitrarily strong
as the super-inflation phase evolves is a universal feature of Veneziano's 
string cosmology in all spacetime dimensions~\cite{universal}. 
This indicates that 
classical description of string cosmology breaks down and that quantum 
effects modify evolution of the two classical branches significantly. 
One may then wonder if quantum back reaction can wash out curvature 
and string coupling divergences completely that a transition to FRW phase 
becomes possible. In this Letter, we show that branch change, hence, 
graceful exit is indeed possible after quantum back reaction is taken into 
account. 

To address quantum back reaction in a controlled approximation 
we consider an exactly soluble two dimensional string compactification model.
The model is described by the same action as the dilaton gravity of 
Callan, Harvey, Giddings and Strominger (CGHS)~\cite{cghs}
\begin{equation}
S_0 = \int {d^2 x \over 2 \pi} {\sqrt {-g}} [e^{-2 \phi}
(R + 4 (\nabla \phi)^2 -  4 \Lambda)
-{1 \over 2} (\nabla {\vec f})^2 ]
\label{2daction}
\end{equation}
where $\phi, \Lambda ( \ge 0)$ and ${\vec f}$ denote dilaton, 
cosmological constant (central charge deficit) and free $N$-component 
Ramond-Ramond scalar field. We first show that counterparts of 
Veneziano's two branches with accelerating and decelerating expansions 
are also present in two dimensions.
Equations of motion derived from Eq.(\ref{2daction}) are  
\begin{eqnarray}
&& e^{-2 \phi} \big[ 4 \nabla_\mu \nabla_\nu \phi + g_{\mu \nu} R \big] 
= \nabla_\mu {\vec f} \cdot \nabla_\nu {\vec f}
-{1 \over 2} g_{\mu \nu} (\nabla {\vec f})^2;
\nonumber \\
&& R = 4 (\nabla \phi)^2 - 4 \nabla^2 \phi + 4 \Lambda,
\nonumber \\
&& \nabla^2 {\vec f} = 0.
\label{eqnmotion}
\end{eqnarray}
In conformal gauge
\begin{equation}
ds^2 = - e^{2 \rho} d x_+ d x_-; \hskip0.2cm x_\pm = t \pm x, 
\hskip0.2cm \partial_\pm = {1 \over 2} (\cdot \pm \prime),
\end{equation}
the equations of motion Eq.(\ref{eqnmotion}) can be derived from  
the following gauge fixed action
\begin{eqnarray}
S_0 = \int {d^2 x \over 2 \pi} \Big[ e^{-2 \phi } \Big( &4& \partial_+ 
\partial_- \rho -8 \partial_+ \phi \partial_- \phi - 2 \Lambda e^{2 \rho} 
\Big) 
\nonumber \\
&+& \partial_+ {\vec f} \cdot 
\partial_- {\vec f} \Big]
\label{gfaction}
\end{eqnarray}
supplemented with constraint equations
\begin{equation}
T_{\pm\pm}= {1 \over 2} (\partial_\pm {\vec f})^2
+ e^{-2 \phi} \big( 4 \partial_\pm \rho \partial_\pm \phi
-2 \partial_\pm^2 \phi \big) = 0.
\label{constr}
\end{equation}
That the model is classically exactly soluble is seen by changing the 
field variables to~\cite{bilalcallan,rst} 
\begin{equation}
\Phi = e^{-2 \phi}, \hskip1cm \Sigma = 2 \kappa (\phi - \rho)
\label{fieldvar}
\end{equation}
where $\kappa$ is a parameter introduced to keep track of quantum loop 
expansions. 
Then Eqs.(\ref{gfaction}, \ref{constr}) become
~\cite{bilalcallan}
\begin{eqnarray}
S = \int {d^2 x \over \pi } \Big[
&&  {1 \over 2 \kappa} \big( \partial_+ \Phi \partial_- \Sigma
+\partial_- \Phi \partial_+ \Sigma \big) - \Lambda e^{-\Sigma/\kappa} 
\nonumber \\
&+& {1 \over 2} \partial_+ {\vec f} \cdot \partial_- {\vec f} \big]
\label{newaction}
\end{eqnarray}
and
\begin{equation}
T_{\pm\pm} = {1 \over 2} (\partial_\pm {\vec f})^2 
+ \partial_\pm^2 \Phi + {1 \over \kappa} \partial_\pm \Phi \partial_\pm \Sigma
= 0
\label{newconstr}
\end{equation}
respectively. In the classical limit $\kappa \rightarrow 0$, the theory
has a useful $\Sigma$-field symmetry~\cite{rst}
\begin{equation}
\Phi \rightarrow \Phi +  \epsilon_\Sigma, \hskip1cm \Sigma \rightarrow \Sigma.
\label{phisymm}
\end{equation}
In what follows we restrict to the case of vanishing cosmological constant 
$\Lambda = 0$. In this case there is an additional $\Phi$-field symmetry 
\begin{equation}
\Sigma \rightarrow \Sigma + \epsilon_\Phi, \hskip1cm \Phi \rightarrow \Phi.
\label{sigmasymm}
\end{equation}
Thus the theory consists entirely of $(N+2)$ free scalar fields and is
exactly soluble.
The most general homogeneous vacuum solution of equations of motion
\begin{equation}
\ddot \Phi = \ddot \Sigma = 0 ; \hskip1cm {\vec f} = {\rm constant} 
\label{eom}
\end{equation}
is given by
\begin{eqnarray}
-{1 \over 2 \kappa} \Sigma & := & (\rho - \phi) = Q_\Sigma t + A,
\nonumber \\
\Phi & := & \,\,\, e^{-2 \phi} \,\,\, = Q_\Phi t + B
\end{eqnarray}
where $Q_\Sigma, Q_\Phi, A,B$ are integration constants specified by 
initial conditions. In particular $Q_\Sigma, Q_\Phi$ denote conserved 
charges associated with the two classical symmetries Eqs.(\ref{phisymm}), 
(\ref{sigmasymm}). Imposing the constraint Eq.(\ref{newconstr})
\begin{equation} 
{\ddot \Phi} + {1 \over \kappa} {\dot \Phi} {\dot \Sigma} 
= 2 Q_\Phi Q_\Sigma = 0,
\end{equation}
cosmological solutions are classified into two distinct branches 
depending on whether each conserved charge vanishes or not:
$Q_\Sigma = 0, Q_\Phi \ne 0$ or $Q_\Sigma \ne 0, Q_\Phi = 0$.

The first branch having  $Q_\Sigma = 0, Q_\Phi \ne 0$ may be expressed as
\begin{equation}
\rho = \phi + \log 2 {\tilde M} ; \hskip1cm e^{-2\phi} =  - 8 {\tilde M} t
\label{superinf}
\end{equation}
after suitable shift of dilaton and coordinate-time $t$.
Reality condition for dilaton field $\phi$ then restricts 
$ -\infty < t \le 0$. For this interval, by introducing comoving cosmic time
$\tau := - (- 2 {\tilde M} t)^{1/2}$ that ranges over
$ -\infty < \tau \le 0$, the spacetime metric may be 
expressed as 
\begin{equation}
(ds)^2 = - {{\tilde M}^2 \over -2 {\tilde M} t} [dt^2 - dx^2] = 
- [d \tau^2 - ({{\tilde M} \over - \tau})^2 dx^2].
\label{superinfmetric}
\end{equation}
The comoving scale factor $a(\tau) = {\tilde M} /(-\tau)$ shows 
that this branch describes super-inflationary evolution.
Dilaton in this branch evolves as $\phi = - \log (-2\tau)$.

The second $Q_\Sigma \ne 0, Q_\Phi = 0$ branch is similarly expressed as
\begin{equation}
\rho = \phi + M t ; \hskip1cm e^{-2\phi} = M^{-2}.
\label{milne}
\end{equation}
The spacetime metric describes an expanding universe
\begin{equation}
(ds)^2 = - M^2 e^{2Mt} [dt^2 - dx^2] = - [ d\tau^2 - (M \tau)^2 dx^2]
\label{milnemetric}
\end{equation}
where comoving time $\tau: = \exp Mt$ ranges over $0  \le \tau < \infty$.
Universe expanding linearly in time is known as Miln\'e universe.
The Miln\'e universe is a flat two-dimensional counterpart of the FRW-type
universe. Dilaton in Miln\'e universe branch is frozen to a constant value.

It is straightforward to recognize that the above two branches are 
mapped each other by scale-factor duality as in higher dimensions.
In fact the two branches were known as anisotropic universes of
Bianchi I type~\cite{gsv}. Is it then possible to complete the super-inflation 
and reheat to the Miln\'e universe by matching the two branches at $\tau = 0$?
That this is not possible is immediately recognized from singular behavior 
of curvature scalar and string coupling constant in the super-inflation
branch
\begin{equation}
g_{\rm st} = {1 \over - 2 \tau} \rightarrow \infty, \hskip0.5cm   
R = ({2 \over - \tau})^2 \rightarrow \infty
\label{singularities}
\end{equation}
in contrast to regular $g_{\rm st}$ = finite, $R=0$ in the Miln\'e branch.
Thus any cosmological vacua belonging to the super-inflation branch end up 
with infinite curvature and string coupling and cannot escape out of that 
branch forever. Nature of the difficulty is inherently the same as in higher 
dimensional string inflationary cosmology, hence, may be viewed as 
two-dimensional version of the aforementioned \sl graceful exit problem \rm.

On a closer look, however, it is clear that one has to go beyond classical 
description to string cosmology. Since curvature $R$ and string coupling 
$g_{\rm st}$ in the super-inflationary branch diverge as $\tau \rightarrow 
0^-$, quantum corrections cannot be neglected. It may even be that the 
quantum back reactions are sufficiently strong enough to wash out the
classical divergences Eq.(\ref{singularities}) that a smooth transition
to the Miln\'e branch becomes possible.
One-loop quantum correction is provided by conformal anomaly of 
$N$-component matter, reparametrization ghosts, dilaton and conformal
modes. 
After adding local, covariant counterterms to retain the two classical
symmetries Eqs.(\ref{phisymm}),(\ref{sigmasymm}) for exact 
solvability~\cite{bilalcallan,rst}, the quantum effective action is given by
\begin{equation}
S_{\rm eff} = S_0 - {\kappa \over 2} \int {d^2 x \over 2 \pi} {\sqrt - g}
\big[ R {1 \over \Box } R + 2 \phi R \big]
\label{effaction}
\end{equation}
where $\kappa = (N-24)/24$. The second term in Eq.(\ref{effaction}) is 
a local, covariant counterterm and is added to retain the two classical
symmetries Eqs.(\ref{phisymm}),(\ref{sigmasymm}), hence, the exact
solvability. This is most easily seen from the fact that the effective 
action $S_{\rm eff}$ has exactly the same form as the classical one
Eq.(\ref{newaction}) in terms of quantum corrected fields 
\begin{equation}
\Sigma = 2 \kappa (\phi - \rho); \hskip1cm
\Phi = e^{-2 \phi }  + \kappa \rho.
\label{quantumfieldvariables}
\end{equation}
Structure of the kinetic term then suggests that quantum theory defined 
in terms of $S_{\rm eff}$ introduces a new expansion 
parameter~\cite{russotseytlin}
\begin{equation}
g_{\rm eff}^2 = 1 / |2e^{-2\phi} - \kappa| .
\label{geff}
\end{equation}
The theory in fact corresponds to an exactly soluble 
conformal field theory~\cite{bilalcallan}.
The constraint equations Eq.(\ref{newconstr}) are also modified to
\begin{equation}
\partial_\pm^2 \Phi + {1 \over \kappa} \partial_\pm \Phi \partial_\pm 
\Sigma + (\partial_\pm {\vec f})^2 
- \kappa \big({1 \over 2} \partial_\pm \Sigma + t_\pm (x^\pm) \big)
= 0.
\label{quantumconstr}
\end{equation}
The first integrals $t_\pm(x^\pm)$ come from nonlocality of conformal anomaly
and are determined by boundary conditions. We set $t_\pm = 0$ by demanding
the correct classical behavior for flat Minkowski spacetime vacuum with a 
constant string coupling.

Since the equations of motion are the same as classical ones Eq.(\ref{eom}), 
the most general homogeneous \sl quantum \rm solutions are again given by  
\begin{eqnarray}
{1 \over 2 |\kappa|} {\overline \psi} := (\rho - \phi) &=& Q_\Sigma t + A 
\nonumber \\
\Phi := e^{-2 \phi} - |\kappa| \rho &=& Q_\Phi t + B.
\label{qusol}
\end{eqnarray}
For reasons that will become clear immediately, we have restricted 
$\kappa < 0$, viz, $N < 24$. We also note that the kinetic term of 
$(\rho, \phi)$ is non-degenerate only for this choice.
Moreover the constraint equation Eq.(\ref{quantumconstr})
\begin{eqnarray}
\kappa t_\pm &=& 0
\nonumber \\
&=& {\ddot \Phi} +{|\kappa| \over 2} {\ddot \Sigma} - {1 \over |\kappa|} 
{\dot \Phi} {\dot \Sigma} = Q_\Phi Q_\Sigma,
\label{quconstr2}
\end{eqnarray}
implies that there are again two solution branches. 
We now show that the two branches are not distinct but one and
the identical, in sharp contrast to the classical situation.
Both branches describe a universe evolving from super-inflation phase
to Miln\'e universe phase without encountering physical singularities. 

The first quantum branch that corresponds classically to the 
super-inflationary branch is given by
\begin{equation}
\rho = \phi + \log 2 {\tilde M}  ; \hskip1cm
e^{-2 \phi} - |\kappa| \rho = - 8 {\tilde M} t.
\end{equation}
Combining the two, we have
\begin{eqnarray}
&& e^{-2 \rho} - {|\kappa| \over 4 {\tilde M}^2} \rho = 
- {2  t \over {\tilde M} }
\nonumber \\
&& 
e^{-2 \phi} - |\kappa| \phi = - 8 {\tilde M} t + |\kappa| \log 2{\tilde M}.
\label{qurel}
\end{eqnarray}
To appreciate significance of $\kappa < 0$ choice we note that the 
left hand sides of Eq.(\ref{qurel}) are monotonic functions ranging over 
$(-\infty, + \infty)$.
It also implies that real-valued $\rho, \phi$ evolution is possible
for $-\infty < t < + \infty$, hence, extends beyond the classical 
evolution range $-\infty < t \le 0$. Since $\phi$ is a monotonic function 
of $t$, one may take string coupling $g_{\rm st} = e^\phi$ as a convenient 
built-in clock.

At asymptotic past infinity $ t \approx - \infty$, 
\begin{eqnarray}
(ds)^2 &\rightarrow& -\Big({{\tilde M}^2 \over -2 {\tilde M} t}\Big) 
[dt^2 - dx^2] = - [d \tau^2 - ({{\tilde M} \over -\tau})^2 dx^2],
\nonumber \\
\phi &\rightarrow& -\log (-2 \tau),
\end{eqnarray}
hence, the universe starts with classical super-inflationary phase.
It is straightforward to recognize that accelerating expansion 
$\dot a(\tau) > 0$ and $\ddot a(\tau) > 0$ is maintained for $\tau \ll 0$. 
On the other hand, at $t \rightarrow + \infty$, 
\begin{eqnarray}
(ds)^2 &\rightarrow & - e^{16 {\tilde M} t / |\kappa|} 
[dt^2 - dx^2] = - [d \tau^2 - ({8 {\tilde M} \over |\kappa|} \tau)^2 dx^2],
\nonumber \\
\phi &\rightarrow & \log \tau,
\end{eqnarray}
where $\tau \approx (|\kappa| / 8 {\tilde M}) 
\exp (8 {\tilde M} t / |\kappa|)$,
thus behaves as Miln\'e universe with a linear dilaton.
Again it is straightforward to recognize that $ \dot a(\tau)  
> 0$ but $\ddot a(\tau) < 0$, viz, decelerating expansion for
$ \tau \gg 0$.

Given that $\rho, \phi$ evolves continuously it is clear that initially 
super-inflationary accelerating expansion is changed into decelerating 
expansion at late time and approaches asymptotically to Miln\'e universe.
More insight to this may be gained from the behavior of scalar curvature
as a function of coupling parameters $g_{\rm st}$ or $g_{\rm eff}$:
\begin{eqnarray}
R &:&= 2 e^{-2 \rho} (d^2 \rho / dt^2)
\nonumber \\
&=& 16{g_{\rm st}^2 \over (1 + |\kappa| g_{\rm st}^2/2)^3 }
\nonumber \\
&=& 32 g_{\rm eff}^2 (1 - |\kappa| g_{\rm eff}^2 )^2.
\end{eqnarray}
It shows that the curvature vanishes at past/future infinity 
$g_{\rm st} \rightarrow 0/\infty$ and attain the maximum near 
`transition epoch' $ \tau \approx 0$ where $g_{\rm st} \approx 1$. 
The last line shows that $g_{\rm eff}$ reaches the maximum value $\sim 
1 / |\kappa|$ at the transition epoch.
We conclude that the classical super-inflation branch has made 
successful `graceful exit' to the Miln\'e branch with the aid of 
quantum back reaction.

Similarly, the second quantum branch that corresponds classically to the 
Miln\'e universe becomes
\begin{equation}
\rho = \phi + Mt ; \hskip1cm e^{-2 \phi} - |\kappa|  \rho = M^{-2},
\end{equation}
hence,
\begin{eqnarray}
e^{-2 \rho} - |\kappa| e^{-2 Mt} \rho &=& M^{-2} e^{-2Mt}
\nonumber \\
e^{-2 \phi} - |\kappa| \phi &=& |\kappa| M t + M^{-2}.
\end{eqnarray}
At $ t \rightarrow - \infty$, the branch behaves 
\begin{eqnarray}
(ds)^2 &\rightarrow& - \exp\Big(2(M^{-2} - e^{2 M t})/|\kappa|\Big) 
[dt^2 - dx^2] \nonumber \\
&=& -[d\tau^2 - a(\tau)^2 dx^2]; \hskip0.8cm 
a(\tau) \approx \exp(e^{2M\tau}),
\nonumber \\
\phi &\rightarrow & - M \tau,
\end{eqnarray}
where $ \tau \approx \int^t dt' \exp (-e^{2Mt'}/|\kappa|) \rightarrow 
- \infty $.
It is straightforward to see that initially Minkowski flat universe 
with a linear dilaton enters quickly into inflationary branch 
$\dot a(\tau) > 0$ and $\ddot a(\tau) > 0$.
At $ t \rightarrow + \infty$, 
\begin{eqnarray}
(ds)^2 &\rightarrow&  -{1 \over |\kappa| Mt} e^{2Mt} [dt^2 - dx^2]
\nonumber \\
&=& -[d\tau^2 - a(\tau)^2 dx^2]; \hskip0.8cm 
a(\tau) \approx {\tau \over (|\kappa| \log \tau)^{1/2}},
\nonumber \\
\phi &\rightarrow& -{1 \over 2} \log (\log \tau),
\end{eqnarray}
hence, the universe exhibits decelerating expansion evolution 
$\dot a(\tau) > 0, \ddot a(\tau) < 0$. 
Asymptotically at $\tau \rightarrow + \infty$, the 
universe approaches a Miln\'e universe with approximately constant
dilaton. Scalar curvature expressed as a funtion of $g_{\rm st}$, 
$g_{\rm eff}$ is given by
\begin{eqnarray}
R &=& C^2 \exp (-{2 \over |\kappa|} {1 \over g_{\rm st}^2})
{g_{\rm st}^4 \over ( 1 + |\kappa| g_{\rm st}^2 / 2)^3}
\nonumber \\
&=& (2C)^2 \exp \big(-{1 \over |\kappa| g_{\rm eff}^2} + 1 \big)
g_{\rm eff}^4 (1 - |\kappa| g_{\rm eff}^2 )
\end{eqnarray}
where $C:= |\kappa| M \exp (1/|\kappa| M)$.
The $g_{\rm st}$ decreases monotonically from strong to weak coupling with 
$\tau$. The curvature scalar vanishes at asymptotic past and future infinities
$g_{\rm st} \rightarrow 0, \infty$ but approaches a positive maximum value
near transition epoch $\tau \approx 0$, $g_{\rm st} = {\cal O}(1)$. 
In terms of effective coupling the curvature vanishes at 
$\tau \rightarrow \pm \infty$ where $g_{\rm eff} \rightarrow 0$ but 
becomes strong $g_{\rm eff} \rightarrow {\cal O} (1 / |\kappa|)$ near 
transition epoch.
It is interesting to note that the curvature scalar depends on both 
couplings in a suggestive form of nonperturbative effect associated with
particle pair production.
We conclude that the second quantum branch also shows \sl graceful exit \rm 
evolution of inflationary branch into Miln\'e universe branch much the same 
manner as the first quantum branch. 

One common aspect of both quantum branches is that the string coupling
$g_{\rm st} = e^\phi$ evolves monotonically: either from weak to strong for 
the first branch or from strong to weak coupling for the second. 
Normally this indicates breakdown of string perturbation theory.
On the other hand recent advent of string-string dualities~\cite{duality} 
may suggest alternative interpretation: branch interpolation between 
strong and weak string coupling is equivalent to an interpolation between 
two weakly coupled phases of dual string pairs~\cite{stringduality} such as 
type-I and heterotic and type-II/M-theory and
heterotic strings. String duality then allows that $g_{\rm st} \rightarrow
0$ both at asymptotic past and future infinity but for different perturbative 
string theories. 
We also note that the effective coupling $g_{\rm eff}$ behaved in a similar
way as the string-duality-applied $g_{\rm st}$. 
Interestingly the transition epoch is precisely where both $g_{\rm eff}$
and string-duality-applied $g_{\rm st}$ is of order unity, hence, 
perturbation theory in all means should break down or the least well-behaved. 
Evidently this is where quantum back reaction is most pronounced.

In summary we have investigated effects of quantum back reaction in
string cosmology. In exactly solvable two-dimensional model we have found
that the effect is profound: classically disjoint super-inflation and
Miln\'e branches have become smoothly connected at quantum level. 
The consequence is that initially super-inflation phase with accelerating
expansion gets retarded enough that exits gracefully to Miln\'e universe
with decelerating expansion. Thus a successful string inflationary cosmology 
is realized. Whether a similar mechanism is possible for higher dimensions
is an interesting question and is under current investigation.

S.J.R. thanks M. Gasperini and G. Veneziano for many helpful discussions.
This work was supported in part by U.S.NSF-KOSEF Bilateral Grant, 
KRF Nondirected Research Grant and International Collaboration Grant,
KOSEF Purpose-Oriented Research Grant and SRC Program Grant,
and Ministry of Education BSRI 94-2418.

\end{document}